\documentclass{aa} % if the references are not structured 
\usepackage{graphicx}
\usepackage{txfonts}
\usepackage{array}
\usepackage{enumitem}
\newcolumntype{P}[1]{>{\centering\arraybackslash}p{#1}}
\usepackage{subcaption}
\usepackage[export]{adjustbox}
\usepackage{siunitx}
\usepackage{multirow}
\usepackage{natbib}
\usepackage{hyperref}
\hypersetup{
     colorlinks   = true,
     citecolor    = blue
}

\defcitealias{Rivera_2018}{R18}
\defcitealias{Wijnands_Parikh_2019}{Wijnands \& Parikh (2019)}
\defcitealias{Parikh_Wijnands_2019}{Parikh \& Wijnands (2019)}

\begin{document} 

   \title{Near-ultraviolet detections of four dwarf nova candidates in the globular cluster 47 Tucanae}

   \author{David Modiano
    \and Aastha S. Parikh
    \and Rudy Wijnands}

   \institute{Anton Pannekoek Institute for Astronomy, University of Amsterdam, Postbus 94249, 1090 GE Amsterdam, The Netherlands\\
   \email{d.modiano@uva.nl}}

   \date{}
 
  \abstract
  {We investigate near-ultraviolet (NUV) variability in the Galactic globular cluster (GC) 47 Tucanae (47 Tuc). This work was undertaken within the GC sub-project of the Transient UV Objects project, a programme which aims to find and study transient and strongly variable UV sources. Globular clusters are ideal targets for transient searches because of their high stellar densities and large populations of variable systems. Using all 75 archival observations of 47 Tuc obtained with the UV/optical telescope (UVOT) aboard the Neil Gehrels \textit{Swift} observatory with the \textit{uvm2} filter, we searched for UV variability using a specialised pipeline which utilises difference image analysis. We found four clear transients, hereafter SW1--4, with positions consistent with those of known cataclysmic variables (CVs) or CV candidates identified previously using Hubble Space Telescope observations. All four sources exhibit significant outbursts, likely brightening by several orders of magnitude. Based on the inferred outburst properties and the association with known CVs, we tentatively identify the UV transients as CV-dwarf novae (DNe). Two DNe have been previously observed in 47 Tuc: V2, which has a position consistent with that of SW4; and AKO 9, which was not in outburst during any of the UVOT observations. We thus increase the known number of DNe in 47 Tuc to 5 and the total number of detected DNe in all Galactic GCs combined from 14 to 17. We discuss our results in the context of the apparent scarcity of DNe in GCs. We suggest that the likely cause is observational biases, such as limited sensitivity due to the high background from unresolved stars in the GC and limited angular resolution of the telescopes used. We additionally detected one strongly variable source in 47 Tuc, which could be identified as the known RR Lyrae star HV 810. We found its period to have significantly increased with respect to that measured from data taken in 1988.}


   \maketitle

\section{Introduction}

Extreme variability in astrophysical sources has been explored extensively in most wavelength regimes. In addition, more recent multi-messenger searches have been performed using gravitational waves \citep{Abbott_2016} and neutrinos \citep{IceCube_2018}. A notable exception, however, is the ultraviolet (UV), which has been used mostly for follow-up observations of transient sources discovered using other means. Systematic large-scale searches for serendipitous transient and highly variable sources in the UV have not been performed (some blind searches have been undertaken in the past but they were limited in scope; e.g. \citealp{Welsh_2005,Wheatley_2008,Gezari_2013}). This is in part a consequence of the opacity of the Earth's atmosphere to most UV radiation, which inhibits ground-based UV transient-searching facilities, and the high costs of space-based missions. Nonetheless, currently operational space-based UV facilities can be very useful in detecting and studying transients in the UV because they often have repeating observations of the same fields. Typically, these telescopes have relatively small fields of view (FOVs), so only a small number of objects can be studied per field. However, because of the large number of fields studied, and through selection of fields of high stellar density (such as galaxy centres and Galactic open and globular star clusters), observations carried out by these facilities can still provide large numbers of possible transients. The Transient UV Objects (TUVO) project, within which the work for this paper was undertaken, aims to exploit this idea in order to discover and characterise transients and highly variable systems in the UV. For a full description of the scientific aims, observing strategy, and operation of the TUVO project see \citetalias{Wijnands_Parikh_2019} and \citetalias{Parikh_Wijnands_2019}.\

The TUVO-Globular project is a TUVO sub-project which targets Galactic globular clusters (GCs). Globular clusters are excellent laboratories for studying variability in a wide range of different types of objects. The high stellar densities of GCs, up to $10^{6}$ $M_\odot$ pc$^{-3}$ (see e.g. \citealp{Knigge_2008}), result in frequent stellar interactions, often producing enhanced populations of variable systems such as cataclysmic variables (CVs), X-ray binaries, and millisecond radio pulsars \citep{Hut_1992,Knigge_2008,Campos_2018, vandenBerg_2019}. In addition, GCs are known to harbour large populations of variable stars, in particular RR Lyrae variables (for a UV study of RR Lyrae in GCs, see \citealp{Siegel_2015}). Several tens to hundreds of other types of variables such as SX Phoenicis, Cepheid, and Red/Mira variables have also been identified in GCs (for a catalogue of variable stars observed in GCs, see \citealp{Clement_2001} and \citealp{Clement_2017}). Galactic GCs are opportune environments in which to probe these objects because relatively accurate estimates of their ages, metallicities, and distances are often known. Although observations of GCs suffer from significant stellar crowding, this is reduced in the UV since stellar populations in GCs are typically dominated by old, low-mass (<1$M_\odot$) main-sequence stars, which are known to emit weakly in the UV, thus reducing the overall background emission from unresolved cluster stars.\

One of the prime targets for our TUVO-Globular project are outbursts from CVs (for reviews on CVs see \citealp{Smith_2006,Giovannelli_2008}). Cataclysmic variables are semi-detached binary systems in which a white dwarf (WD) primary accretes mass from a secondary donor star via Roche lobe overflow. In most of these systems, matter infalling towards the primary forms an accretion disc, although the disc may be truncated by the magnetic field of a WD, if present. Thermal instabilities in the disc cause repetitive outbursts known as dwarf novae (DNe). This phenomenon can be explained in general terms using the disc instability model (DIM; see \citealp{Lasota_2001} or \citealp{Hameury_2019} for reviews of this model, and \citealp{Dubus_2018} for recent tests of the model for DNe). During a DN, the system increases in brightness by 2--6 magnitudes in the optical (see e.g. \citealp{Dobrotka_2006,Belloni_2019}). Outbursts have durations of days to weeks and recurrence times of weeks to decades. The emission from outbursts from accreting WDs peaks in the near-UV (NUV; \citealp{Giovannelli_2008}). Despite this, the UV has not been used extensively to study these types of sources. The NUV emission originates from viscous heating within the accretion disc and potentially from the WD surface, but the lack of detailed and systematic studies of DNe in the UV hinders our understanding of the processes involved.\

The presence of CVs and DNe in GCs has been studied by several groups in the last $\sim$25 years (see e.g. \citealp{Grindlay_1995,Grindlay_2001,Shara_1996,Knigge_2002,Knigge_2008,Bond_2005,Kaluzny_2005,Shara_2005,Dobrotka_2006,Pietrukowicz_2008,Servillat_2011,Belloni_2016, Belloni_2019}). However, investigations have struggled to reconcile observations with theory, in particular with respect to the discrepancy between the predicted and detected numbers of CVs and DNe in GCs. Theoretical predictions estimate that massive GCs should harbour >100 CVs and at least half of them should exhibit DN outbursts \citep{Downes_2001,Dobrotka_2006,Knigge_2011,Belloni_2016}, if not 99\% of them \citep{Belloni_2019}. We should therefore expect, given the number of observations of GCs, to have identified a large number of such events in GCs. However, until now, only 14 GC DNe have been confirmed \citep{Pietrukowicz_2008,Kaluzny_2009,Servillat_2011}. Some additional variable sources in GCs have been suggested to be DNe (e.g. sources W51 and W56 in \citealp{Edmonds_2003}), although the nature of these remains to be confirmed.\ 
The discrepancy between theory and observations has motivated speculation regarding the underlying reasons. There have been several suggestions that some physical characteristics of GC CVs may be different to CVs in the field such that GC DNe are inherently rare events \citep{Belloni_2016,Servillat_2011,Pietrukowicz_2008}. These proposals have included the following: 1) GC CV accretion rates during quiescence are lower than those of field CVs, resulting in less frequent outbursts (see e.g. \citealp{Edmonds_2003}, who indeed found low accretion rates for quiescent CVs in 47 Tuc), although it is unknown why GC CVs should have relatively low accretion rates. 2) The unstable regions of the disc which cause DN outbursts may not be present if the disc is truncated by strong magnetic fields, preventing DNe, which requires that WDs in GC CVs have stronger magnetic fields than
WDs in field CVs (\citealp{Lasota_2001}). 3) The low metallicity of the accreted material may affect the ionisation rate of the accretion disc;  typical GC populations are old and thus have low metal content. Since DNe are triggered by a critical ionisation rate, there may be a metallicity-outburst dependence \citep{Gammie_1998}. 4) A combination of 1) and 2), i.e. \citet{Dobrotka_2006} investigated several possibilities and concluded that the most feasible way of creating non-outbursting CVs in GCs is for the WDs to have moderately strong magnetic fields as well as low accretion rates. The reasons why GC CVs should differ from field CVs in this way, however, is still unclear. It is also possible that current theory overestimates the numbers of GC CVs, for example if these high stellar density environments are more efficient at dynamically destroying than creating binaries. If indeed fewer CVs are created than previously thought, we should logically expect fewer DNe.\

Despite these extensive investigations on possible physical GC CV properties, it has been argued that observational selection effects are the primary cause of the absence of observed DN events (e.g. \citealp{Servillat_2011,Curtin_2015,Belloni_2016}). In this scenario, the numbers of GC CVs are predicted correctly and have most properties in common with field CVs so they indeed exhibit DN outbursts, but owing to both intrinsic characteristics (e.g. recurrence and duration times of the outbursts, peak luminosities compared with bright cluster background), and observational constraints (e.g. the cadence and limiting magnitudes of the observations, and angular resolution of the telescope compared to the crowdedness of the GCs), they are missed. To determine the exact causes of the apparent lack of DNe in GCs in our Galaxy, further observations of these systems need to be undertaken. As detailed above, using UV facilities can significantly aid in this endeavour owing to the intrinsic UV brightness of the objects and the reduced cluster background, which can make DNe easier to detect in the UV compared with the optical. However this also depends on the angular resolution of the telescope used, which could be better in the optical than in the UV.\

47 Tucanae (47 Tuc) is a GC with a total mass of $6.45\pm0.4\times10^{5} M_\odot$ at a distance of $4.69\pm0.04$ kpc, with a low extinction of $E(B-V)=0.04$, low metallicity (Fe/H=-0.76), and high stellar interaction rate (see Table 1 of \citealp{Rivera_2018}, hereafter \citetalias{Rivera_2018}, and references therein for a detailed description of the characteristics of 47 Tuc). Many CVs and candidate CVs have been observed in this cluster (e.g. \citetalias{Rivera_2018}), and this makes it an excellent target, even relative to other GCs, to investigate the UV variability properties of these types of objects (i.e. to detect DNe outbursts from them). Until now, only two confirmed DNe have been discovered in 47 Tuc, known in the literature as V2 \citep{Paresce_1994} and AKO 9 \citep{Knigge_2003}.\

In this work we present the first results of the TUVO sub-project TUVO-Globular, with the aim of studying strongly variable UV sources located in 47 Tuc. In Section 2, we summarise the observations used in this paper and the image processing undertaken to detect sources. In Section 3, we discuss the UV transients we discovered and their candidacy as DNe. We also report on their UV characteristics (e.g. peak brightness and duration/recurrence timescales) and their likely associations with known sources in 47 Tuc. In Section 4, we discuss the implications of these findings in the context of CVs and DNe in GCs.

\section{Observations and data reduction}

\subsection{Observations}
For our study we used archival data from the Ultraviolet/Optical Telescope (UVOT) on board the Neil Gehrels \textit{Swift} observatory \citetext{Gehrels et al., 2004; hereafter referred to as \textit{Swift}}. The UVOT has a 17$'$ x 17$'$ FOV and is equipped with three optical filters (\textit{u}, \textit{b}, \textit{v}) and three UV filters (\textit{uvw1}, \textit{uvm2}, \textit{uvw2}), as well as a magnifier, two grisms, a clear white filter, and a blocked filter (see \citealp{Roming_2005,Breeveld_2010,Breeveld_2011} for a detailed description of the UVOT). The principal scientific goal of the UVOT is to study the optical and UV afterglows of gamma-ray bursts (see e.g. \citealp{Page_2019}). Nonetheless, the broad wavelength range of the telescope ($\sim$1600-8000 \si{\angstrom}; \citealp{Breeveld_2010}) and its many repeating observations of the same fields (due to its rapid slew capabilities) also allows for effective transient searches, despite its small FOV (see \citetalias{Parikh_Wijnands_2019} for a discussion about the UVOT observing strategy and its usefulness in searching for UV transients).\

All data were obtained from the \textit{Swift} data archive at NASA/GSFC\footnote{https://swift.gsfc.nasa.gov/archive/}. The UVOT was used to observe 47 Tuc 76 times over the $\sim$5.5-year period from 11 Feb 2013 to 21 Aug 2017  with cadence between one day and several months; currently no additional observations of
47 Tuc have been performed with \textit{Swift}. All observations (with observation identification numbers or ObsIDs 000497540[01-23] and 000841190[01-56]) were performed in imaging mode and during all but one observations (ObsID 00084119001), the \textit{uvm2} filter (which has a central wavelength of 2246 \si{\angstrom}; \citealp{Poole_2008,Breeveld_2011}) was used. In this one outlying observation the \textit{uvw1} and \textit{uvw2} filters were used. Since we cannot search for variability without a comparison data set for these two filters, this observation was not used further to search for transients and we only used the data obtained with the \textit{uvm2} filter. However, we used the \textit{uvw1} image of this observation to aid with obtaining accurate astrometric solutions to our images (see Section 3.1). An advantage of the \textit{uvm2} filter for UV studies is that it has the smallest optical contamination (the `red leak') of the three UV filters (see e.g. \citealp{Siegel_2015}). However, the 2175 \si{\angstrom} interstellar absorption bump is strongest at the wavelength of the \textit{uvm2} filter \citep{Irvine2011}, thereby reducing our sensitivity to UV sources compared with the two other UV filters. Yet this effect is not very severe because of the low extinction towards 47 Tuc. To monitor targets throughout a given day, \textit{Swift} observations are typically composed of multiple exposures (`snapshots') separated by up to a few hours and available as extensions in the observation files. We extracted every snapshot for the 75 used observations of 47 Tuc and analysed each one individually (resulting in a total of 245 images to process). Exposure times per snapshot ranged between $\sim$0.03-2.3 ks.

\subsection{Image processing}

To detect and analyse highly variable sources, we used the specialised pipeline \texttt{TUVOSpipe} (for a detailed description of the pipeline, see \citetalias{Parikh_Wijnands_2019}). In this section we give a brief description of the main parts of \texttt{TUVOSpipe} and how we implemented it to search for transients in the 47 Tuc data. Both within the pipeline and during our additional data processing (see Section 3.1), we used the tools available in HEAsoft (version 6.25\footnote{https://heasarc.gsfc.nasa.gov/docs/software.html}) with CALDB version 20170922.

Part I (\texttt{TUVOSdownload}) of \texttt{TUVOSpipe} obtains the data from the \textit{Swift} archive (or quicklook page). The raw (`Level I') images have been pre-reduced with the standard UVOT reduction pipeline\footnote{https://swift.gsfc.nasa.gov/quicklook/swift\_process\_overview.html}, producing `Level II' images which are astrometrically solved (with an uncertainty of a few arcseconds), flatfielded, and with bad pixels identified. These are the products obtained by \texttt{TUVOSdownload}.\

Part II (\texttt{TUVOSsearch}) uses difference imaging to search for transients, whereby a template image is subtracted from the science images. Sources with different brightness in the science images with respect to the template image are visible as residuals in the difference images. To perform the subtraction, \texttt{TUVOSsearch} uses the High Order Transform PSF and Template Subtraction (\texttt{hotpants}) package \citep{Becker_2015}, which uses the image subtraction algorithm of \citet{Alard_1998}. The process requires as input one science image and one template image, which must be taken with the same filter, aligned, and of equal size. Additional processing is therefore required because of the uncertainty in the astrometric solutions of UVOT images and the pointing of \textit{Swift}. The downloaded UVOT Level II images can be misaligned by up to several arcseconds. Therefore, before performing each subtraction, \texttt{TUVOSsearch} aligns the science image to the template image. Additionally, \textit{Swift} has a median pointing accuracy of 1.2 arcminutes\footnote{https://swift.gsfc.nasa.gov/proposals/tech\_appd/swiftta\_v14/node23.html}, so observations of the same field cover slightly different regions on the sky. To allow \texttt{hotpants} to be able to process the images properly, \texttt{TUVOSsearch} therefore also crops each science and template image such that they cover exactly the same sky region; this reduces the FOV by up to a few arcminutes, but in our case still includes most of the stars in the GC. The first image (in order of ObsID) is chosen as the template image (ObsID 00049754001; extension 1) from which all other 244 images are subtracted\footnote{We note that choosing a different template image does not affect the results presented in our paper.}.\ 

The image subtraction was found to work suitably well. Although some structure is present and some bright sources leave subtraction artefacts, the difference images are reasonably uniform (see the examples shown in Fig.~\ref{fig:main_sub}). Bright transients (seen as residual sources in the difference images) are clearly visible. The performance of \texttt{hotpants} may be slightly reduced towards the cluster core owing to significant diffuse emission. However, from the difference images it is clear that this is not a significant effect in the case of 47 Tuc, as the cluster core is also well subtracted. Bright transients in the core are indeed clearly visible in the difference images (see bottom panel in Fig.~\ref{fig:main_sub}). The performance of \texttt{hotpants} is also reduced for extremely bright sources which saturated the UVOT detector and thus already have associated artefacts from the UVOT reduction pipeline; only one such foreground source is present in the 47 Tuc data, which is visible in the first, second, and fourth panels of Fig.~\ref{fig:main_sub}.\

\texttt{TUVOSsearch} runs \textit{Source Extractor}\footnote{https://www.astromatic.net/software/sextractor} on every difference image to detect variable sources and produces a file listing the position of each candidate transient. We inspected all images of each candidate to determine which were likely to be real transients and not artefacts of a badly subtracted source. We also examined the light curves produced automatically by Part III of the pipeline, i.e. \texttt{TUVOSanalyse}. However, we note that these default light curves were not useful for this particular study since the pipeline is not designed for fields with strong background due to the diffuse emission produced by, in our case, unresolved stars in 47 Tuc. For each real transient, we undertook further analysis, including the construction of more accurate light curves (see Section 3) using methods specific to these sources which are not implemented generally in the pipeline.\

\section{Analysis and results}

\begin{figure*}[h]
\captionsetup{font=small}
\centering
\begin{adjustbox}{minipage=\linewidth}
\centering
\begin{subfigure}[b]{0.22\linewidth}
    \centering
\includegraphics[width=1\textwidth]{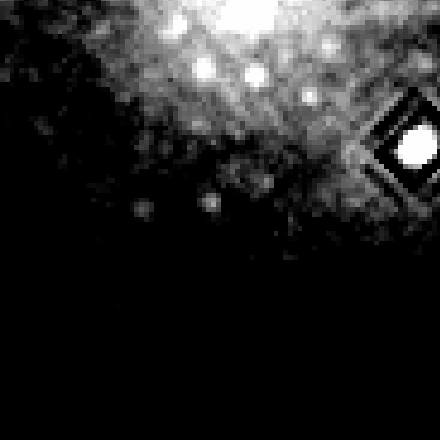}
%\caption*{2013-04-29}
\end{subfigure}\hspace{0.3cm}%
\begin{subfigure}[b]{0.22\linewidth}
    \centering
\includegraphics[width=1\textwidth]{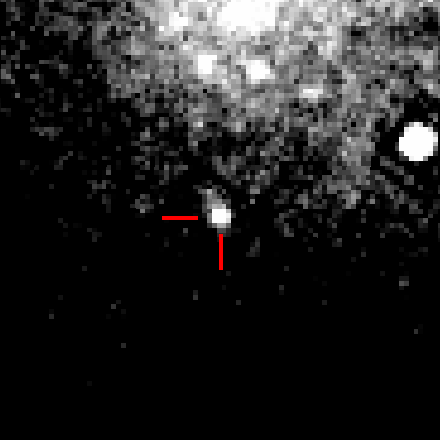}
%\caption*{2013-04-29}
  \end{subfigure}\hspace{0.3cm}%
\begin{subfigure}[b]{0.22\linewidth}
    \centering
\includegraphics[width=1\textwidth]{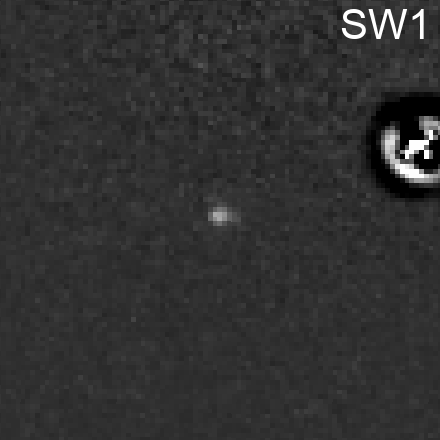}
%\caption*{Difference image}
  \end{subfigure}
  
\vspace{4ex}

 \begin{subfigure}[b]{0.22\linewidth}
    \centering
\includegraphics[width=1\textwidth]{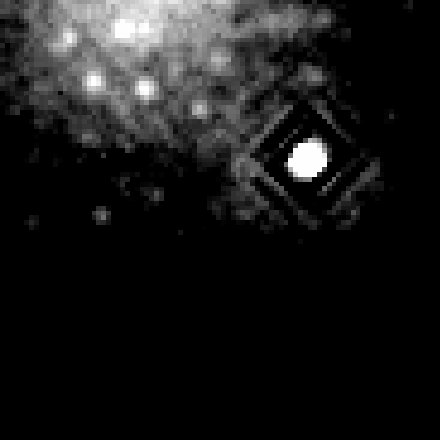}
%\caption*{2013-04-29}
\end{subfigure}\hspace{0.3cm}%
\begin{subfigure}[b]{0.22\linewidth}
    \centering
\includegraphics[width=1\textwidth]{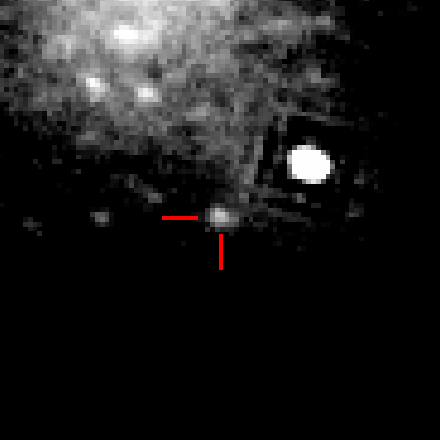}
%\caption*{2013-04-29}
  \end{subfigure}\hspace{0.3cm}%
\begin{subfigure}[b]{0.22\linewidth}
    \centering
\includegraphics[width=1\textwidth]{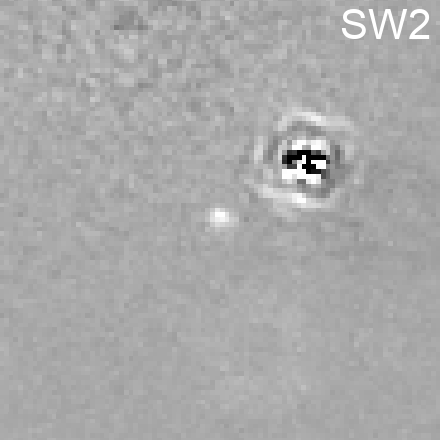}
%\caption*{Difference image}
  \end{subfigure}
  
\vspace{4ex}

 \begin{subfigure}[b]{0.22\linewidth}
    \centering
\includegraphics[width=1\textwidth]{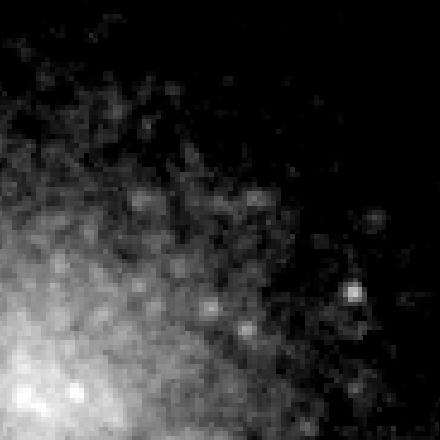}
%\caption*{2013-04-29}
\end{subfigure}\hspace{0.3cm}%
\begin{subfigure}[b]{0.22\linewidth}
    \centering
\includegraphics[width=1\textwidth]{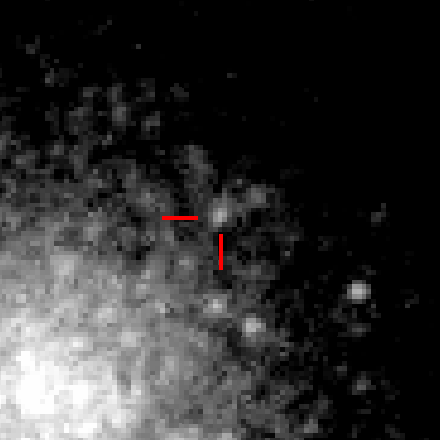}
%\caption*{2013-04-29}
  \end{subfigure}\hspace{0.3cm}%
\begin{subfigure}[b]{0.22\linewidth}
    \centering
\includegraphics[width=1\textwidth]{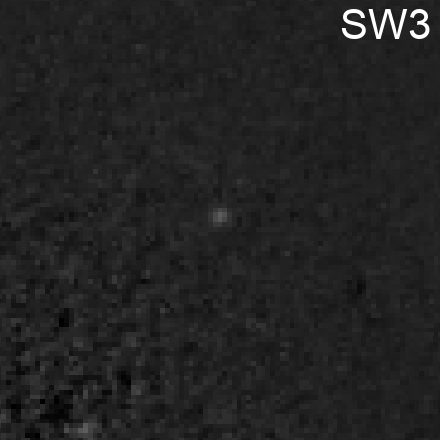}
%\caption*{Difference image}
  \end{subfigure}
  
\vspace{4ex}

 \begin{subfigure}[b]{0.22\linewidth}
    \centering
\includegraphics[width=1\textwidth]{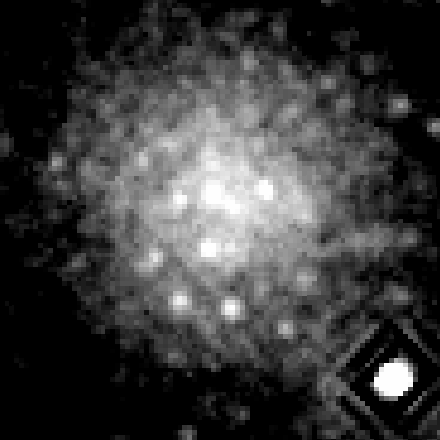}
%\caption*{2013-04-29}
\end{subfigure}\hspace{0.3cm}%
\begin{subfigure}[b]{0.22\linewidth}
    \centering
\includegraphics[width=1\textwidth]{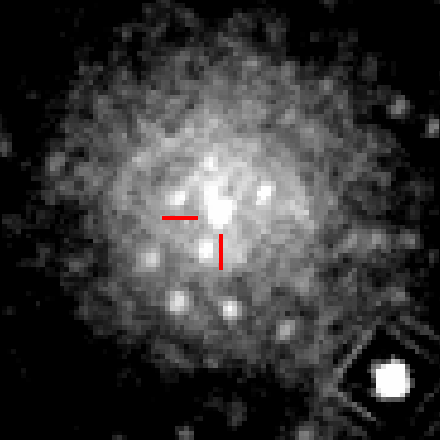}
%\caption*{2013-04-29}
  \end{subfigure}\hspace{0.3cm}%
\begin{subfigure}[b]{0.22\linewidth}
    \centering
\includegraphics[width=1\textwidth]{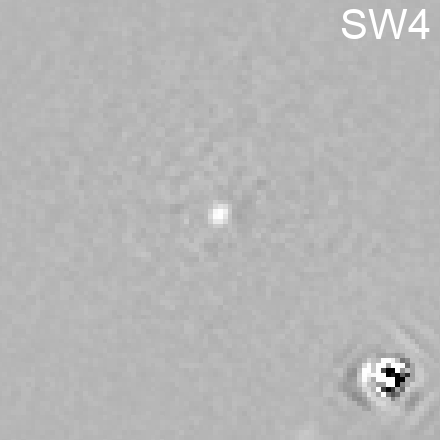}
%\caption*{Difference image}
  \end{subfigure}
\caption{Image subtraction examples for the four transient sources identified in the UVOT data. Left: the template image; centre: sample science images in which enhanced activity of the sources was detected; right: the corresponding difference image. All images are 1.5$'$ x 1.5$'$. North is up and east is to the left. In the science images, the detected transients are indicated by red lines. The template image corresponds to (the first snapshot of) the observation with ObsID 00049754001. The science images correspond to (the first snapshot of) the observations of ObsID 00049754022, 00084119009, 00049754016, and 00084119023, respectively, for SW1--SW4.}
    \label{fig:main_sub}
    \end{adjustbox}
\end{figure*}

Using all available UVOT \textit{uvm2} data of 47 Tuc, we found four transient sources (see Fig.~\ref{fig:main_sub}) and one strongly variable source. For each of the transients, we performed the following analysis to constrain the nature of the objects and their characteristics (Section 3.1). The results of this analysis are then discussed for each source (Section 3.2).

\newpage

\newgeometry{left=1.5cm,bottom=2cm,right=1.5cm,top=2cm}

\renewcommand{\arraystretch}{1.2}

\begin{table*}[!h]
\small
\captionsetup{font=footnotesize}
    \centering
    \begin{tabular}{c|c|c|c|c}
       \textbf{Source Name} & \textbf{Source ID} & \textbf{RA} & \textbf{Dec} & \textbf{Type}\\
       \hline
       \hline
       SWIFT J002407$-$720546 & SW1 & 00:24:06.87 & -72:05:46.8 & Transient\\
       SWIFT J002402$-$720541 & SW2 & 00:24:02.11 & -72:05:41.1 & Transient\\
       SWIFT J002358$-$720413 & SW3 & 00:23:57.72 & -72:04:12.6 & Transient\\
       SWIFT J002406$-$720455 & SW4 & 00:24:05.82 & -72:04:55.2 & Transient\\
       SWIFT J002340$-$720600 & HV 810 & 00:23:40.41 & -72:05:59.7 & Variable\\
    \end{tabular}
    \caption{Names, source IDs as referred to in this paper, coordinates, and classifications as transients or variables for each of our detected sources. The errors on the positions are $\sim$1.2".}
    \label{tab:coords}
\end{table*}

\restoregeometry

\setlist[itemize]{label=\textbullet}

\subsection{Analysis}
We manually constructed light curves using the standard UVOT photometry tool \textit{uvotsource}. This tool performs aperture photometry using a circle (chosen by the user, typically with a radius of 5") centred on the position of the source and gives as output the magnitudes (we used the AB magnitudes) and fluxes of the source, as well as 1$\sigma$ statistical and systematic errors. Systematic errors are derived from errors in the photometric calibration of UVOT (see the instrument manual\footnote{https://swift.gsfc.nasa.gov/analysis/UVOT\_swguide\_v2\_2.pdf} or e.g. \citealp{Breeveld_2011}). The \textit{uvotsource} tool also corrects for coincidence loss of the UVOT, though we note that for the measured magnitudes (see Table~\ref{tab:sum_results}), this effect is of the order of only few percent\footnote{https://swift.gsfc.nasa.gov/analysis/uvot\_digest/coiloss.html}. To obtain accurate source parameters, the background has to be subtracted. Typically a background region is selected close to the source, devoid of bright sources, representative of the underlying background of the target source, and significantly larger than the 5" target source region; \textit{uvotsource} uses the mean pixel value within the background region, so increasing the size reduces bias from Poissonian statistics. Such a region is difficult to find in data containing GCs, since the high stellar densities cause source confusion and blurring. This results in a cluster background in addition to the normal sky background in the UVOT images. At the locations of our sources, the cluster background is much more significant than the sky background, although both backgrounds were corrected for. The sky background can vary significantly both from image to image, for example as a consequence of zodiacal light and Earth shine (see discussion in \citealp{Breeveld_2010}), and across a single image because of internal scattered light within the telescope and detector system (\citealp{Breeveld_2010}). This (relatively small) normal sky background was accounted for as much as possible by providing \textit{uvotsource} with a background region near the edge of the image, far from the GC. This region is therefore dominated by the sky background and not the cluster background. This does not account for the gradient in the background within each image; however this is a small effect, in particular for \textit{uvm2} (see \citealp{Breeveld_2010}). The cluster background also has a strong gradient (dependent on the distance to the centre of the cluster), which makes determining the contribution of the cluster background at our source positions particularly challenging. To correct for this we took advantage of the fact that our sources are transients and thus were only detected in a small fraction of the UVOT images. As a cluster background, we therefore took the mean count rates at our source positions (using the same extraction region as used for the detected transients) in all images in which the source was not detected. The fluxes at our source positions in images when the sources were detected were then corrected by this mean background flux.\

A further benefit of creating light curves manually is that we can stack snapshots within ObsIDs, a feature not yet implemented in \texttt{TUVOSpipe}. The light curves that we built are therefore obtained using one data point per observation to decrease the errors on the magnitudes and fluxes. Light curves with one data point per snapshot were also made, but because of the size of the error bars no useful information could be obtained concerning short timescale (<1 day) variability.\

We examined variability amplitudes and timescales (where possible). We used the maximum observed fluxes with respect to the fluxes from non-detections (i.e. the flux due to the cluster background) to estimate source outburst amplitudes. These estimates are in all cases lower limits, as the observations are unlikely to occur exactly at outburst maximum, and the exact brightness of the sources in quiescence is likely significantly lower than the cluster background at the source location. We also used the light curves to constrain the durations and recurrence times of the outbursts.\

In order to aid identification of the transient sources we found, we compared the coordinates of our sources with known sources identified in 47 Tuc in previous optical and X-ray studies. We therefore needed to determine the positions of our sources as accurately as possible. The automatic UVOT pipeline can astrometrically solve images to $\leq$0.5" \citep{Poole_2008,Breeveld_2010}. But this aspect correction may fail, for example when the UV sky is too different from the optical Digital Sky Survey, and this was the case for all the 47 Tuc data. Therefore, the misalignment between the images was up to several arcseconds (as expected for UVOT images in cases where the aspect correction fails; \citealp{Poole_2008,Breeveld_2010}). We thus obtained accurate astrometric solutions by running our images through astrometry.net. Since astrometry.net uses optical reference images, it could not obtain accurate coordinate solutions for our \textit{uvm2} images. However, it could successfully solve the coordinates for the (redder) \textit{uvw1} image (ObsID 00084119001). We then used the image alignment function from the \textit{scikit-image}\footnote{https://scikit-image.org/docs/dev/api/skimage.feature.html} package to align all our \textit{uvm2} images to the solved \textit{uvw1} image. The mean error on the astrometric solution was calculated using the rms error between the expected and measured positions of the solved astrometry.net image and found to be 1.2". Therefore, we have an error circle of 1.2" for each of our sources. The source coordinates obtained are listed in Table~\ref{tab:coords}. \

Multiple studies, including \citetalias{Rivera_2018}, identified tens of confirmed and candidate CVs in a census of 47 Tuc using Hubble Space Telescope (HST) observations, using optical colours, variability, H$\alpha$ excess, and association with X-ray sources identified in 47 Tuc using the \textit{Chandra X-ray Observatory}. If sources exhibited blue colours, optical variability, and H$\alpha$ excess they were classified as CVs; if sources exhibited only blue colours they were classified as CV candidates; this is the definition used in this paper when we refer to CVs and CV candidates. For each of our UV transients, we determined if it could be associated with an optical/X-ray source from one of these studies. We note that our error circle (as described above) is significantly larger than that of the WFC3/UVIS instrument on HST used by the \citetalias{Rivera_2018} study (1.2" vs <0.1"). This introduces uncertainty in our identification because as a result of the stellar densities of 47 Tuc, there are often multiple sources within 1.2" of the HST source (see e.g. Fig. 3 of \citealp{Edmonds_2003} or Fig. B1-B2 of \citetalias{Rivera_2018}, which show HST images of typical CV candidates in 47 Tuc with FOV $\sim$2" and $\sim$1" across, respectively). This suggests that any source we detect at the location of an HST source may in principle be associated with any of these sources. However, all four of the potential \citetalias{Rivera_2018} CV and CV candidate counterparts to our sources are separated from other CVs and CV candidates in that study by at least a few arcseconds (always >1.2"), indicating that our sources can only be associated with one known CV or CV candidate. We used our measured \textit{uvm2} magnitudes (representing the sources in outburst) and the HST U300 magnitudes of the likely associated sources (representing the sources in quiescence) to infer minimum outburst amplitudes. The U300 filter is the bluest with which CVs and CV candidates in 47 Tuc were identified in \citetalias{Rivera_2018}. Therefore despite being only an estimate of the \textit{uvm2} magnitude of the source in quiescence due to the $\sim$800 \si{\angstrom} difference in central wavelength between the \textit{uvm2} and U300 filter, this gives an indication of the UV outburst amplitude of the sources.\

We estimated the X-ray flux of our sources during their brightest observed outbursts using the simultaneous \textit{Swift} X-Ray Telescope (XRT) observations. For sources for which there are no X-ray detections at the source locations, we used the XRT exposure times and the 3$\sigma$ confidence limit prescriptions from \citet{Gehrels_1986} to determine the count rate upper limits. We used the HEASARC tool Webpimms\footnote{https://heasarc.gsfc.nasa.gov/cgi-bin/Tools/w3pimms/w3pimms.pl} to infer upper limits for the unabsorbed X-ray flux (for the energy range 0.5--10 keV), assuming a spectrum with a power-law index of 2 (as observed by e.g. \citealp{Balman_2015} in DNe) and Galactic or source-specific (if available) column densities $N_{H}$ observed for each source by \citet[see Table 7 in their paper]{Heinke_2005}. See Table~\ref{tab:fluxes_lums} for the upper limits on the X-ray fluxes and corresponding luminosity upper limits of our sources at their brightest outbursts and the details of the parameters used to derive upper limits for each source. We also list the Chandra X-ray (0.5-6 keV) luminosities for the likely quiescent counterparts of our sources as reported in \citet{Heinke_2005}.\

We used the distance of 4.69 $\pm$ 0.04 kpc \citep{Woodley_2012} and the effective filter bandpass of \textit{uvm2} of 533 \AA\ to determine the \textit{uvm2} luminosities for the brightest observed outbursts of each source (see Table~\ref{tab:fluxes_lums}) to compare these values roughly with typical UV luminosities of DNe (see e.g. \citealp{Wheatley_2000} and references therein; \citealp{Ramsay_2010}).\

\subsection{Results}

Five sources displaying strong variability or transient behaviour were found in the difference images: SWIFT J002407$-$720546, SWIFT J002402$-$720541, SWIFT J002358$-$720413, SWIFT J002406$-$720455, and SWIFT J002340$-$720600, hereafter referred to as SW1, SW2, SW3, SW4, and the RR Lyrae HV 810, respectively; the last source can conclusively be identified as this RR Lyrae star (see Section 3.2.6). See Table~\ref{tab:coords} for the names, positions, and IDs of our sources. In Fig.~\ref{fig:main_sub} we show 1.5$'$ x 1.5$'$ stamps of the template image and examples of science and difference images for SW1--4. The \textit{uvm2} fluxes for every ObsID in which each transient was detected are shown in Table~\ref{tab:observations_tab}. The measured peak fluxes and corresponding luminosities in \textit{uvm2} and XRT (upper limits) for the brightest outburst we observed for each transient are given in Table~\ref{tab:fluxes_lums}. The results for SW1--4 are summarised in Table~\ref{tab:sum_results}.

\renewcommand{\arraystretch}{1}

\begin{table}[t]
\small
\captionsetup{font=footnotesize}
    \centering
    \begin{tabular}{P{1cm}|c|c|P{3cm}}
    \textbf{Source} & \textbf{ObsID} & \textbf{Date} & \textbf{Flux} ($\mathbf{10^{-15}\ erg\,s^{-1}\,cm^{-2}}$\,{\bfseries \AA}$\mathbf{^{-1}}$) \\
    \hline
    \hline

    \multirow{6}{4em}{\textbf{SW1}} & 00049754004 & 2013-07-02 & 1.62$\pm$0.18 \\ 
    & 00049754022 & 2013-11-06 & 3.23$\pm$0.25 \\
    & 00084119002 & 2014-11-17 & 0.40$\pm$0.10 \\
    & 00084119010 & 2015-03-05 & 1.38$\pm$0.14 \\
    & 00084119019 & 2016-01-03 & 2.43$\pm$0.19 \\
    & 00084119020 & 2016-01-04 & 2.34$\pm$0.20 \\

    \hline
    
    \multirow{17}{4em}{\textbf{SW2}} & 00049754001 & 2013-02-11 & 1.41$\pm$0.19 \\
    & 00049754011 & 2013-08-17 & 0.75$\pm$0.15 \\
    & 00049754012 & 2013-08-19 & 0.95$\pm$0.16 \\
    & 00049754013 & 2013-08-22 & 1.17$\pm$0.18 \\
    & 00049754014 & 2013-08-24 & 0.86$\pm$0.16 \\
    & 00049754015 & 2013-08-27 & 0.40$\pm$0.18 \\
    & 00084119003 & 2014-11-29 & 0.46$\pm$0.13 \\
    & 00084119006 & 2015-01-07 & 0.72$\pm$0.14 \\
    & 00084119009 & 2015-02-09 & 1.47$\pm$0.19 \\
    & 00084119022 & 2016-01-29 & 1.35$\pm$0.19 \\
    & 00084119023 & 2016-02-03 & 0.90$\pm$0.16 \\
    & 00084119027 & 2016-03-25 & 0.37$\pm$0.14 \\
    & 00084119036 & 2016-07-07 & 0.96$\pm$0.16 \\
    & 00084119043 & 2017-02-26 & 0.51$\pm$0.14 \\
    & 00084119046 & 2017-03-16 & 0.60$\pm$0.14 \\
    & 00084119051 & 2017-05-17 & 0.81$\pm$0.15 \\
    & 00084119053 & 2017-06-18 & 0.76$\pm$0.15 \\
    
    \hline
    
    \multirow{3}{4em}{\textbf{SW3}} & 00049754016 & 2013-09-03 & 0.91$\pm$0.13 \\ 
    & 00049754017 & 2013-09-04 & 0.75$\pm$0.11 \\
    & 00049754018 & 2013-09-11 & 0.58$\pm$0.10 \\
    
    \hline
    
    \multirow{5}{4em}{\textbf{SW4}} & 00084119002 & 2014-11-17 & 2.03$\pm$0.48 \\
    & 00084119023 & 2016-02-03 & 3.29$\pm$0.54 \\
    & 00084119034 & 2016-06-17 & 1.77$\pm$0.46 \\
    & 00084119043 & 2017-02-26 & 1.80$\pm$0.48 \\
    & 00084119044 & 2017-03-01 & 1.58$\pm$0.47 \\
    
    \end{tabular}
    \caption{Details of the observations during which our four UV transients were detected. The \textit{uvm2} fluxes are listed for each detection. The errors on the fluxes are the 1$\sigma$ errors (statistical combined with systematic, the latter representing errors in the photometric calibration of UVOT) as obtained using the \textit{uvotsource} tool. The fluxes are corrected for cluster background flux by subtracting the mean flux at the source position at times when the sources are not detected.}
    \label{tab:observations_tab}
\end{table}

\renewcommand{\arraystretch}{1.2}

\subsubsection{SW1}

\begin{figure}[h!]
\captionsetup{font=footnotesize}
    \centering
    \includegraphics[trim={0 1cm 0 0},clip,width=1\columnwidth]{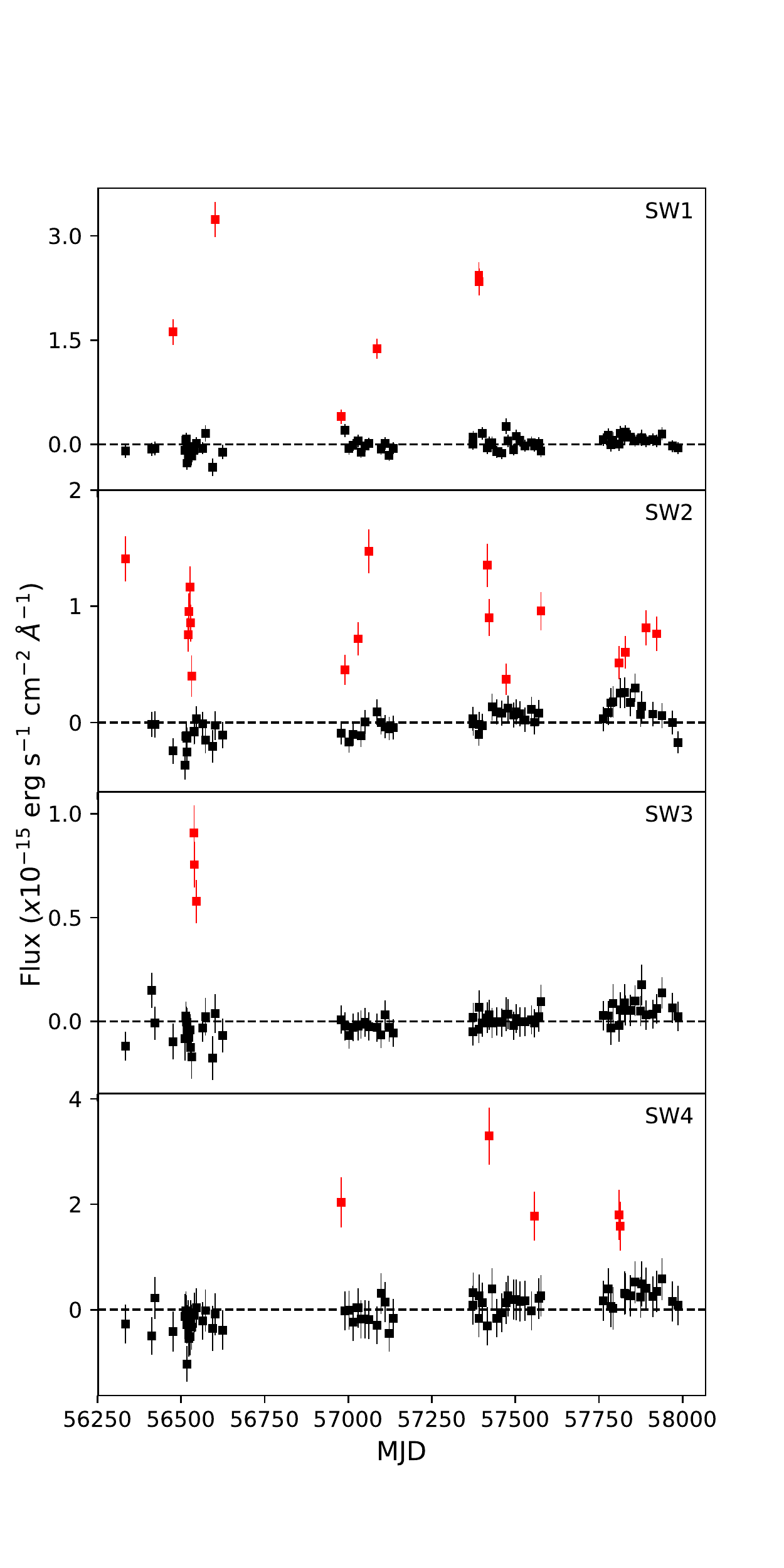}
    \caption{Light curves from \textit{uvm2} for all four UV transients. The red points are detections and the black points are non-detections. The errors bars are the total 1$\sigma$ (statistical and systematic) errors as obtained using the \textit{uvotsource} tool. All data points are corrected for cluster background by subtracting the mean cluster flux at the source position when the sources are not detected. The black dashed lines are the averages of the black points after correction (thus fixed at a flux of 0).}
    \label{fig:lightcurves}
\end{figure}

This source was detected in six of the UVOT observations at a position $\sim$0.89$'$ from the cluster centre (see Fig.~\ref{fig:main_sub}, top row, Table~\ref{tab:observations_tab}, and Fig.~\ref{fig:lightcurves}, top panel). No source was visible above the cluster background at the same location in all other UVOT images. To check whether a deeper image may reveal the source while not in full outburst, we stacked all observations in which the source was not detected using the HEASARC tool \textit{fappend}\footnote{https://heasarc.gsfc.nasa.gov/lheasoft/ftools/fhelp/fappend.txt} (to append images) and the UVOT tool \textit{uvotimsum}\footnote{https://heasarc.gsfc.nasa.gov/lheasoft/ftools/headas/uvotimsum.html} (to sum the images). This resulted in a deep image; however no source was visible at the location of SW1. The mean cluster background \textit{uvm2} magnitude when the source was not exhibiting enhanced activity (determined using \textit{uvotsource} at the source location) was $17.6\pm0.1$, indicating that the source in quiescence was fainter than this value. The source was detected twice during its fifth outburst and once during each other outburst (see Fig.~\ref{fig:lightcurves} top panel). The maximum observed brightness exhibited by SW1 was during its second outburst, when it was detected at a \textit{uvm2} magnitude of $16.6\pm0.1$.

Owing to the sampling of the data, there were typically gaps of at least one week (and frequently a few weeks) between observations in which the source was detected and preceding/following observations in which the source was not detected. However, the fifth outburst was detected in observations 00084119019 and 00084119020, which were performed one day apart (see Fig.~\ref{fig:lightcurves}, top panel). From the data, we infer that the outburst durations were always <2 weeks. The times between successive observed outbursts were 127 days, 376 days, 108 days, and 304 days. Because of the data gaps in the coverage of 47 Tuc, entire outbursts may have been missed, so these values are upper limits on the outburst recurrence times (see Fig.~\ref{fig:lightcurves}, top panel).\

The position of SW1 is consistent with the location of a CV identified by \citetalias{Rivera_2018} and \citet[W25 in their papers]{Edmonds_2003}. However, the source has never been observed in outburst. \citetalias{Rivera_2018} reported a U300 magnitude of 20.7 for W25. Based on the observed \textit{uvm2} brightness of this source in outburst, this indicates a lower limit for the UV magnitude brightening from 20.7 to 16.6, thus $\gtrsim$4.1 (for its brightest observed outburst).\ 

\subsubsection{SW2}

Our second transient SW2 was detected in the difference images corresponding to 17 different UVOT observations at $\sim$0.87$'$ from the cluster centre (see Fig.~\ref{fig:main_sub}, second row, Table~\ref{tab:observations_tab}, and Fig.~\ref{fig:lightcurves}, second panel). The outbursts of this source recur on relatively short ($\sim$weeks) timescales (see Fig.~\ref{fig:lightcurves}, second panel). Stacking images of non-detections, as was done for SW1, again did not reveal any source above the cluster background. The mean cluster background magnitude when the source was not detected was $17.8\pm0.1$, so the source was fainter than this when not in outburst. The source was observed in outburst 12 times, with five detections during the second outburst, two detections during the sixth outburst, and one detection during each other outburst. At the maximum brightness we observed for SW2 (during the fifth outburst; see Fig.~\ref{fig:lightcurves}, second panel), we measured a \textit{uvm2} magnitude of $17.1\pm0.1$. 

\begin{figure}[h!]
\captionsetup{font=footnotesize}
    \centering
    \includegraphics[width=1\columnwidth]{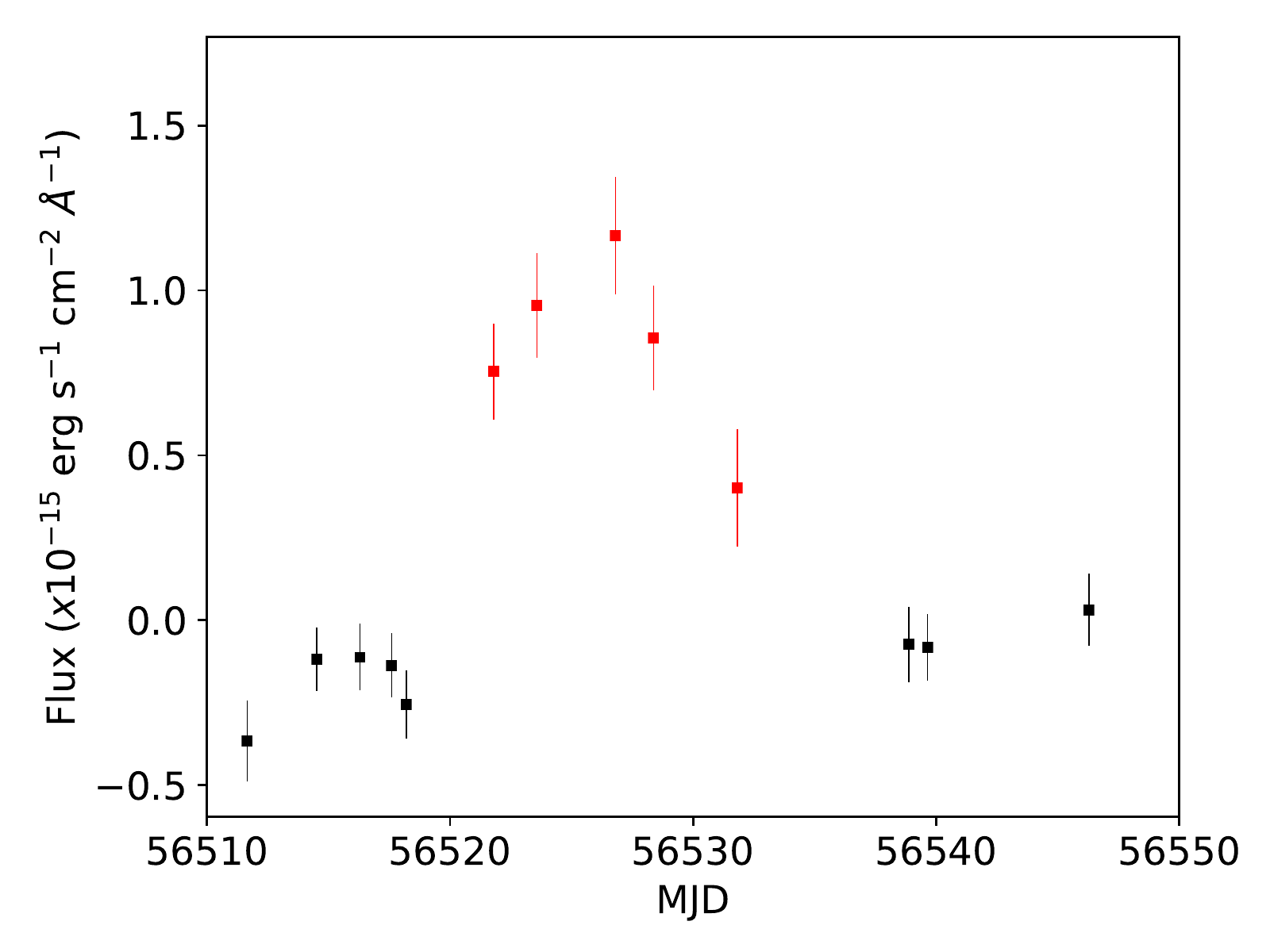}
    \caption{Zoom of the \textit{uvm2} light curve of SW2 during its second observed outburst in the UVOT data, showing that the source exhibited a rise and decay in brightness over a 10 day period, with a maximum duration of the outburst of 20 days. The red points show detections and the black points show non-detections.}
    \label{fig:SW2_zoom2}
\end{figure}

Because of the large number of detected outbursts and the sampling during certain outbursts, both duration and recurrence timescales can be better constrained for SW2 compared to SW1. Although again typical observations adjacent to outburst observations were separated in time by over one week, some non-detections occurred just a few days before or after a detection, providing more accurate estimates of outburst durations. The second outburst visible in the light curve at MJD$\sim$56530 (see Fig.~\ref{fig:lightcurves}, second panel) provides the best time series because both a rise and a decay are clearly visible, suggesting an outburst duration of at least 10 days but not longer than 20 days (as shown more clearly in the zoomed in version of the light curve of this outburst in Fig.~\ref{fig:SW2_zoom2}). The weaker constraints we infer from the other outbursts of this source are consistent with this range. This includes the final four detections (red points in the light curves), which are all separate outbursts. Because of the inhomogenous cadence
and large gaps in the sampling, the recurrence time is not easily
determined. However, the times between separate, successive observed outbursts were typically tens of days. Longer separations between outbursts, of >100 days, occur only occasionally, suggesting that the longer measured times are likely due to the large gaps in the data and that a best estimate for the recurrence time is indeed tens of days (see Fig.~\ref{fig:lightcurves}, second panel).

In their works, \citetalias{Rivera_2018} and \citet{Edmonds_2003} identified a CV (W56 in their papers) at a location that lies within the UVOT error circle of our source SW2, implying a high likelihood that our source corresponds to this known source exhibiting an outburst. The \citetalias{Rivera_2018} authors reported a U300 magnitude of W56 of 20.3, suggesting a UV magnitude brightening from 20.3 to 17.1, thus $\gtrsim$3.2. \citet{Edmonds_2003} detected large amplitude variations for this source, but they did not conclusively attribute this to DN outbursts.\ 

We noticed an apparently unusual rise in background flux at the source position of SW2 during the fourth data period (see Fig.~\ref{fig:lightcurves}, second panel). After visually inspecting the corresponding images, we determined that these elevated points are likely due to the variable background from unresolved stars and potentially artefacts of the bright nearby source (see Fig.~\ref{fig:main_sub}, second row). None of these points represent a significant detection of the source.

\subsubsection{SW3}

\begin{figure}[h!]
\captionsetup{font=footnotesize}
    \centering
    \includegraphics[width=1\columnwidth]{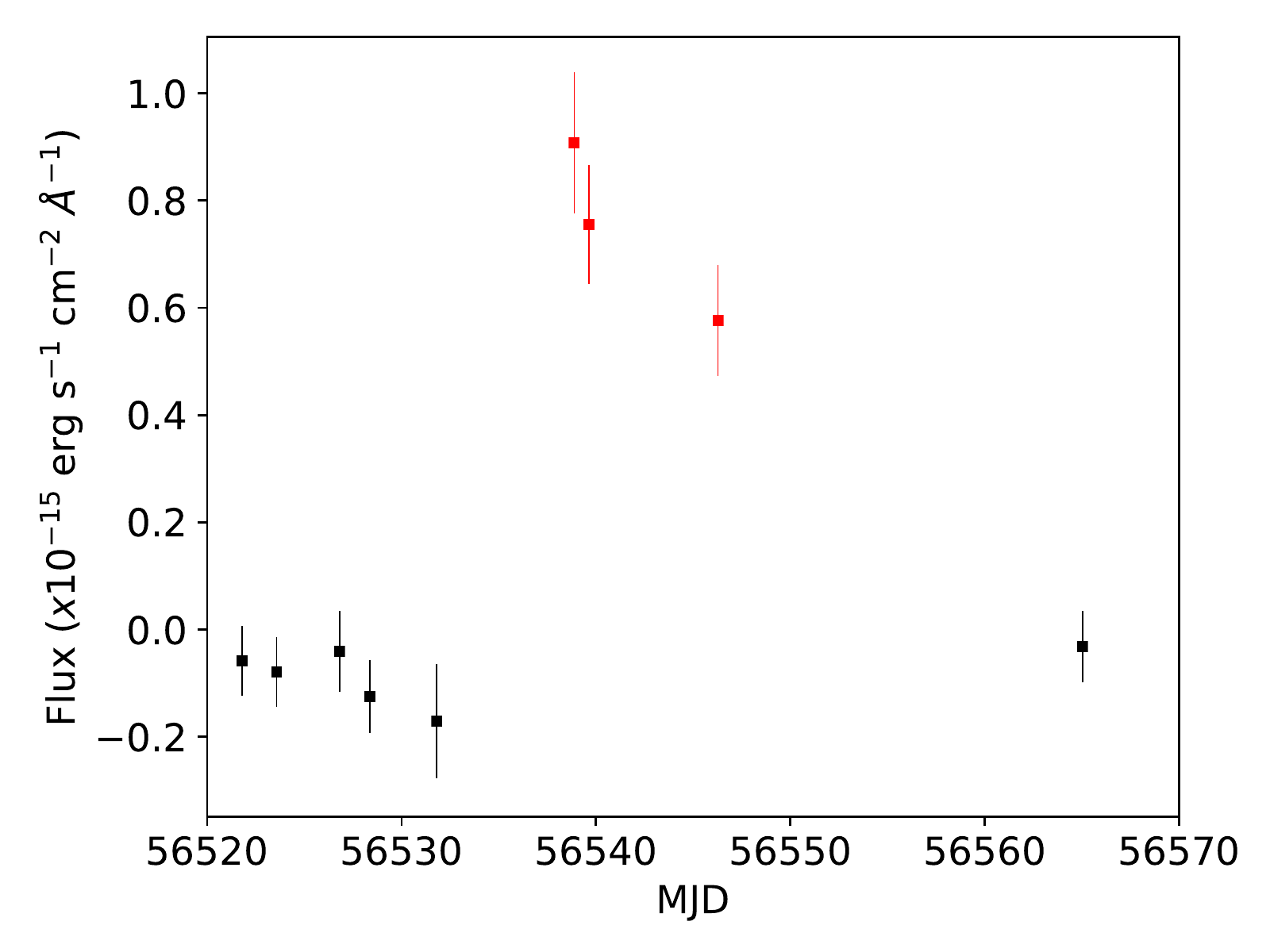}
    \caption{Zoom of the \textit{uvm2} light curve of SW3 during its only observed outburst in the UVOT data, showing that the outburst lasted at most 30 days and that the rise time was only a few days. Owing to the gaps in the data, the decay timescales could not be constrained to better than <20 days. The red points show detections and the black points represent non-detections.}
    \label{fig:SW3_zoom2}
\end{figure}

SW3 was only detected in three (consecutive) UVOT observations over the course of 8 days at a position $\sim$0.89$'$ from the cluster centre (see Fig.~\ref{fig:main_sub}, third row, Table~\ref{tab:observations_tab}, and Fig.~\ref{fig:lightcurves}, third panel). These detections likely represent enhanced activity of the source during a single outburst (see Fig.~\ref{fig:lightcurves}, third panel; and Fig.~\ref{fig:SW3_zoom2}). Using the same method as for SW1 and SW2, we stacked all images in which the source was not detected and found no source at the position of SW3 in the resulting deep image. The mean cluster background \textit{uvm2} magnitude when the source was not in outburst was $18.1\pm0.1$, indicating that the source in quiescence was dimmer than this. The maximum observed flux (the first of the three SW3 detections; see Fig.~\ref{fig:lightcurves}, third panel) corresponded to a \textit{uvm2} magnitude of $17.5\pm0.1$.

Three observations took place during the decay stage of the outburst - the source is seen to decrease in brightness over 8 days (see zoomed in version of the light curve of SW3 in Fig.~\ref{fig:SW3_zoom2}). From preceding and subsequent observations in which SW3 was not detected, we infer a rough duration of the outburst of more than 8 days but less than 30 days. Immediately prior to the outburst observations, several closely spaced observations were taken over roughly 30 days in which no outburst was detected; this gives a lower limit of $\gtrsim$30 days for the recurrence time. No other similar instance occurred in the observations whereby we had consecutive data points each separated by <8 days and spanning more than 30 days combined. Therefore, the best constraint we can establish for the recurrence time of SW3 is indeed $\gtrsim$30 days.

The UVOT error circle of our source SW3 is consistent with a CV candidate identified by \citetalias{Rivera_2018} and \citet[W324 in their papers]{Edmonds_2003}. These authors reported a U300 magnitude of 21.9 of this source. If W324 is indeed the counterpart of SW3, this implies that the UV magnitude of SW3 brightened from 21.9 to 17.5, indicating a brightening on the magnitude scale of $\gtrsim$4.4.\

\subsubsection{SW4}

This source was detected in five UVOT observations (see Fig.~\ref{fig:main_sub}, bottom row, Table~\ref{tab:observations_tab}, and Fig.~\ref{fig:lightcurves}, bottom panel) at a position very near the centre of the cluster. The last two detections correspond to the same outburst, so we observed this source in four different outbursts; these last two detections are separated by three
days with no other observations in between in which the source was not detected. SW4 was also not detected in quiescence as a consequence of very high cluster background at its location. The mean cluster background \textit{uvm2} magnitude when the source was not detected was $16.2\pm0.1$. This is a very weak constraint on its brightness in quiescence since the cluster background at the location of SW4 is significantly greater than at the location of SW1, SW2, and SW3. The maximum observed flux of SW4 (observed during its second outburst, see Fig.~\ref{fig:lightcurves}, bottom panel) corresponded to a \textit{uvm2} magnitude of $15.8\pm0.1$. 

All non-detection observations immediately prior or subsequent to outburst observations occurred at least one week apart from the outburst, indicating rough outburst durations of <2 weeks. The times between successive outbursts were 443 days, 136 days, and 253 days, giving upper limits for the recurrence times of the outbursts (see Fig.~\ref{fig:lightcurves}, bottom panel).\

Our error circle (1.2") of SW4 is consistent with the location of the known DN V2 (W30 in \citetalias{Rivera_2018}). The U300 magnitude in quiescence of this source is reported at 19.7 (\citetalias{Rivera_2018}), indicating a UV magnitude brightening from 19.7 to 15.8, thus $\gtrsim$3.9.\

\begin{table*}[!t]
\small
\captionsetup{font=footnotesize}
    \centering
    \begin{tabular}{P{1.2cm}|P{3.4cm}|P{2.0cm}|P{1.6cm}|P{3cm}|P{2.2cm}|P{2.0cm}}
        \textbf{Source} & \textbf{$\mathbf{F_{UV}}$\ \ \ \ \ \  ($\mathbf{\times10^{-15}\ erg\ s^{-1}\ cm^{-2}}$\ {\bfseries \AA}$\mathbf{^{-1}}$)} & \textbf{$\mathbf{L_{UV}}$} ($\mathbf{\times10^{33}\ erg\ s^{-1}}$) & \textbf{$\mathbf{N_{H}}$} ($\mathbf{\times10^{20}\ cm^{-2}}$) & \textbf{$\mathbf{F_{X}}$ ($\mathbf{\times10^{-13}\ erg\ s^{-1}\ cm^{-2}}$)} & \textbf{$\mathbf{L_{X}}$} ($\mathbf{\times10^{33}\ erg\ s^{-1}}$) & \textbf{Chandra $\mathbf{L_{X}}$} ($\mathbf{\times10^{30}\ erg\ s^{-1}}$)\\
        \hline \hline
        \textbf{SW1} & $3.23\pm0.25$ & $4.54\pm0.23$ & 6.5 & <$6.5$ & <$1.7$ &  $53.1^{+1.8}_{-1.6}$\\ 
        \textbf{SW2} & $1.47\pm0.19$ & $2.07\pm0.12$ & 10.5 & <$0.7$ & <$0.2$ & $134.2^{+2.7}_{-2.6}$\\ 
        \textbf{SW3} & $0.91\pm0.13$ & $1.27\pm0.08$ & 1.3 & <$4.0$ & <$1.1$ & $0.75^{+0.4}_{-0.2}$\\ 
        \textbf{SW4} & $3.29\pm0.54$ & $4.62\pm0.25$ & - & - & - & $47.7^{+1.7}_{-1.6}$\\ 
    \end{tabular}
    \caption{Detected peak fluxes and corresponding luminosities for the brightest observed UV outburst of each transient in the \textit{uvm2} filter and the X-ray upper limits (0.5-10 keV). The errors on the fluxes are the (statistical combined with systematic) 1$\sigma$ errors as obtained using the \textit{uvotsource} tool. The luminosities are calculated using the distance of 4.69 $\pm$ 0.04 kpc and the effective filter bandwidth of \textit{uvm2} of 533\ \AA, and the errors are derived by propagating the errors on the flux and the distance. The XRT upper limits were determined using the HEASARC tool Webpimms and appropriate estimates of $N_{H}$ column densities and the power law index (see Section 3.1 for details). For SW4, no XRT upper limit could be determined due to contamination at the source location from a very bright nearby X-ray source (namely X9; see Section 4). The Chandra X-ray luminosities are taken from \citet{Heinke_2005} for the likely counterparts to SW1-4, namely W25, W56, W324, and W30, respectively. We note that they correspond to a slightly different energy range (0.5-6 keV vs 0.5-10 keV, which we have used) and were converted to the 4.69 kpc distance used in our paper (\citealp{Heinke_2005} used a distance of 4.85 kpc).}
    \label{tab:fluxes_lums}
\end{table*}

\subsubsection{Non-detection of AKO 9}

In addition to the known DN V2 (which likely is SW4; see Section 3.2.4), there is another previously known DN in 47 Tuc, AKO 9 (W36 in \citetalias{Rivera_2018}). This source was first observed in 1989 \citep{Auriere_1989} and was observed in outburst at least once (\citealp{Minniti_1997}, and see \citealp{Knigge_2003} for an overview of the observed properties of AKO 9). In our difference images, we did not detect any residuals at the position of AKO 9, demonstrating that it was not in outburst during any of the UVOT observations.

\subsubsection{RR Lyrae HV 810}

\begin{figure}[h!]
\captionsetup{font=footnotesize}
    \centering
    \includegraphics[scale=0.5]{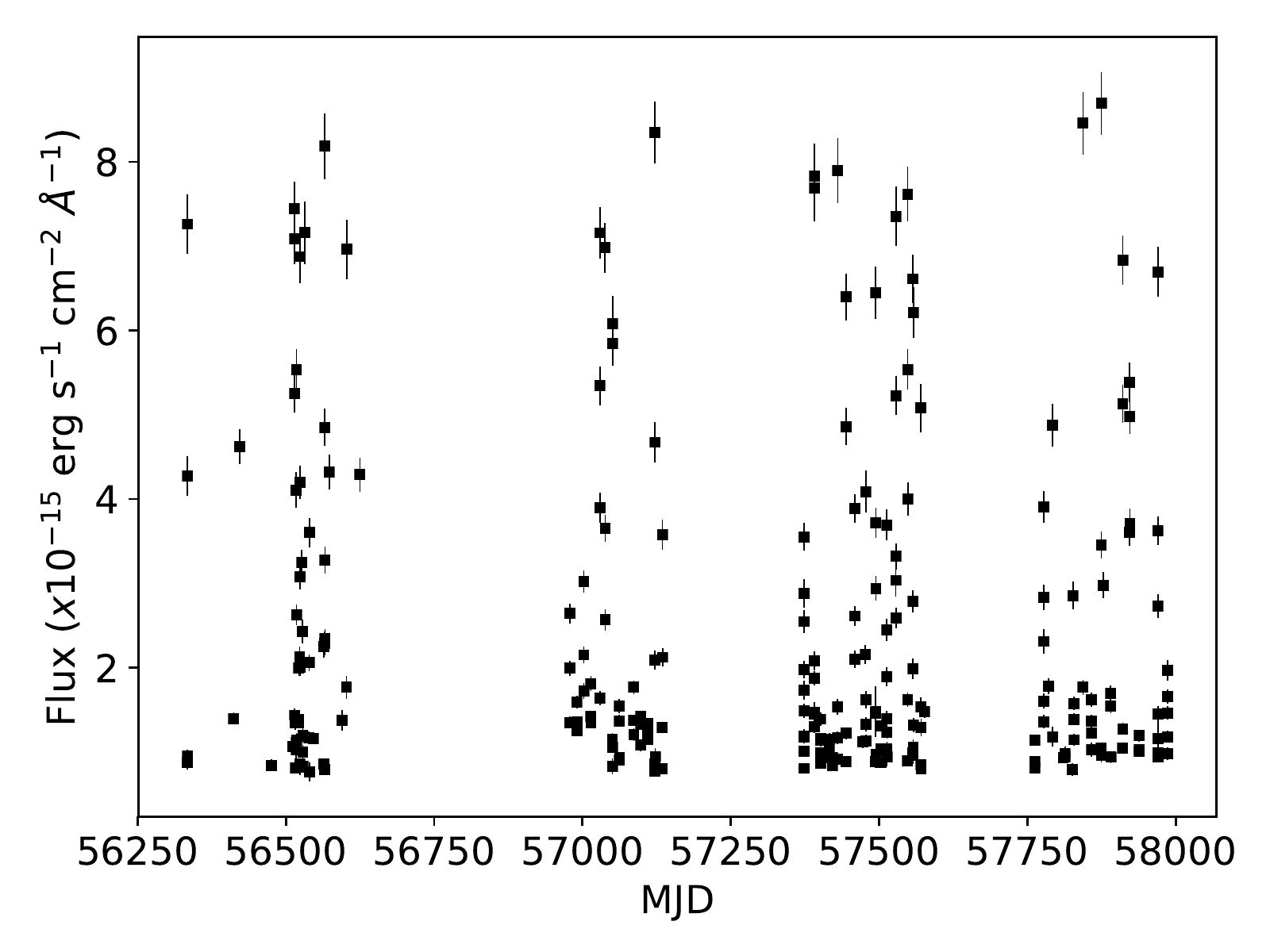}
    \caption{RR Lyrae HV 810 \textit{uvm2} light curve. The times for all data points have been barycentre corrected.}
    \label{fig:rr_lyrae}
\end{figure}

\begin{figure}[h!]
\captionsetup{font=footnotesize}
    \centering
    \includegraphics[scale=0.5]{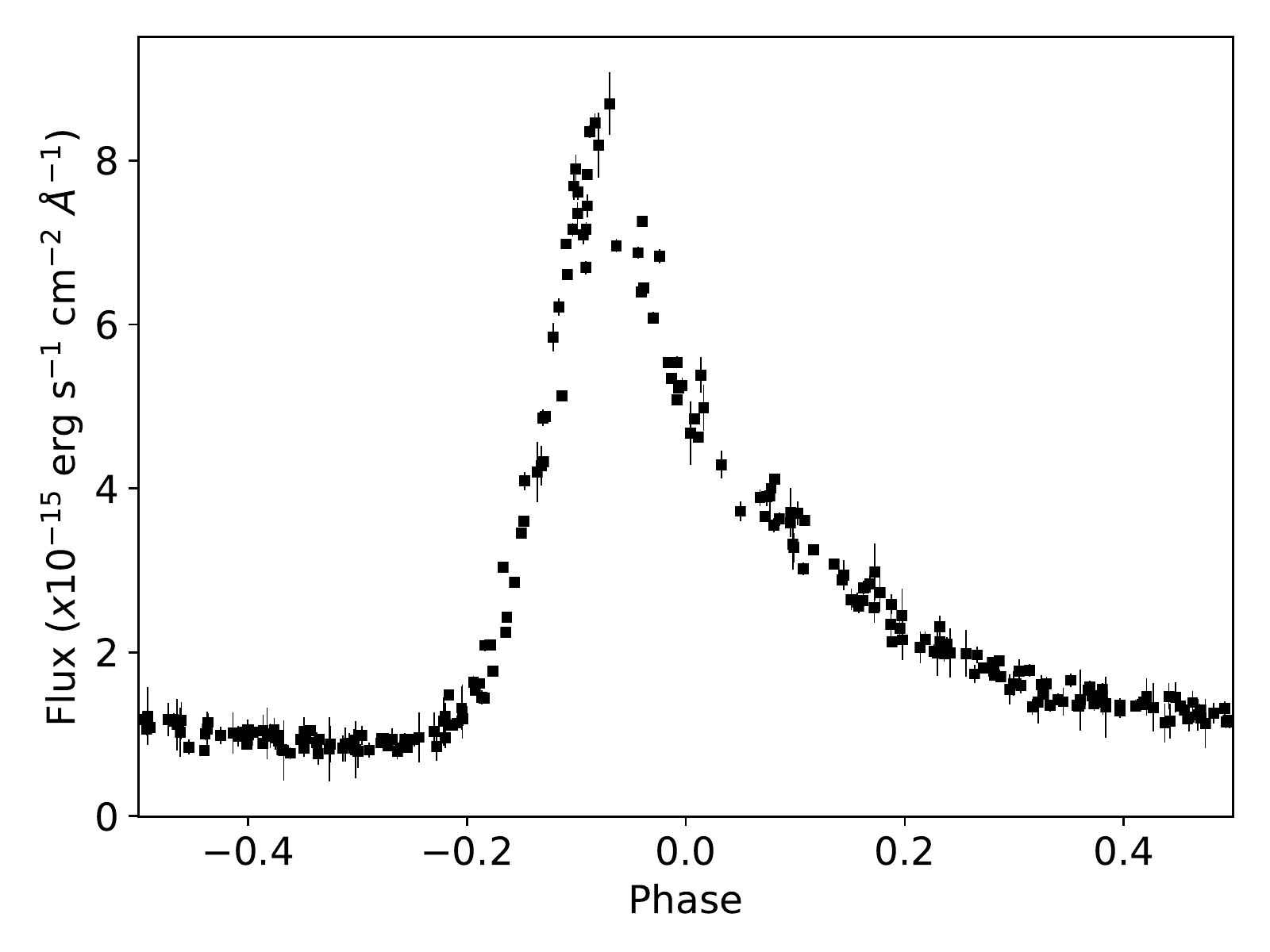}
    \caption{RR Lyrae HV 810 \textit{uvm2} folded light curve using a period of 0.737107 days. The phase is chosen arbitrarily.}
    \label{fig:rr_lyrae_folded}
\end{figure}

We additionally detected bright positive sources in our difference images resulting from the varying UV brightness of the only known RR Lyrae variable in 47 Tuc, HV 810 \citep{Carney_1993}. We applied barycentric corrections to the arrival times of the photons in all images using the HEASARC tool \textit{barycorr}\footnote{https://heasarc.gsfc.nasa.gov/ftools/caldb/help/barycorr.html}. We constructed a barycentre-corrected light curve (shown in Fig.~\ref{fig:rr_lyrae}) and folded it using the previously reported period of HV 810 (0.736852 days) measured from data taken in 1988 \citep{Carney_1993}. We found that the resulting folded light curve did not indicate any periodicity (only scattered points), indicating that the period used was not correct. We used a phase dispersion minimisation analysis (described by \citealp{Stellingwerf}) tool (from the \textit{Python} kit \texttt{pwkit}\footnote{https://pwkit.readthedocs.io/en/latest/science/pwkit-pdm/}) to search our barycentre-corrected light curve for periodic variability and we found a period at 0.737107$\pm$0.000003 days (the error is 1$\sigma$, computed by \texttt{pwkit}). See Fig.~\ref{fig:rr_lyrae_folded} for the light curve folded with this period. This value is larger than the previously reported period by 0.000255$\pm$0.000003 days, indicating a rate of period change for HV 810 over the last few decades of $\beta$ $\sim$10 d Myr$^{-1}$. RR Lyrae are known to undergo period changes, potentially linked with their evolutionary status; for example it has been suggested that RR Lyrae in later stages
of evolution tend to experience higher period change rates, (see \citealp{Jurcsik_2001}). Typical RR Lyrae `period changers' have $\beta$ < 0.1 d Myr$^{-1}$ (see e.g. \citealp{Jurcsik_2001,Smith_2013,ArellanoFerro_2016,ArellanoFerro_2018}), which is much smaller than the period change we determine for HV 810. However, a small number of RR Lyrae have been observed with very large period change rates of 1 < $\beta$ < 10 d Myr$^{-1}$ \citep{Jurcsik_2001}. These were stars with long periods ($\sim$0.5 d) and low mean absolute magnitudes ($M_{V}<1.5$, thus very bright stars), suggested to be in later stages of evolution. HV 810 is also a bright ($M_{V}=1.07$), long-period, highly evolved star \citep{Carney_1993,Storm_1994}. Thus, the large period change implied by our observations is consistent with those found for other, similar RR Lyrae stars.\ 

\citet{Carney_1993} measured optical variability for HV 810 and found changes on the magnitude scale of 1.07 (V) and 1.35 (B), while we measure a \textit{uvm2} amplitude of 2.8 magnitudes. Our results are therefore consistent with the significantly enhanced pulsation amplitudes in the UV (with respect to optical) observed from other RR Lyrae variables (see e.g. \citealp{Siegel_2015}).

\section{Discussion}

\begin{table*}[!t]
\small
\captionsetup{font=footnotesize}
    \centering
    \begin{tabular}{P{1.3cm}|P{2cm}|P{2.2cm}|P{2cm}|P{1.5cm}|P{2.0cm}|P{2cm}|P{1.5cm}}
        \hline
        \hline
        \textbf{Source ID} & \textbf{Distance from cluster centre ($'$)} & \textbf{Brightest observed \textit{uvm2} magnitude} & \textbf{Observed \textit{uvm2} outburst amplitude (magnitudes)} & \textbf{U300 quiescence magnitude} & \textbf{Approximate UV outburst amplitude (magnitudes)} & \textbf{Optical CV or CV candidate counterpart} & \textbf{Previously observed in outburst} \\
        \hline 
        \hline
        SW1 & 0.89 & 16.57$\pm$0.09 & >1.1 & 20.702(8) & >4.1 & W25 & No\\
        SW2 & 0.87 & 17.09$\pm$0.11 & >0.7 & 20.33(2) & >3.2 & W56 & Possibly\\
        SW3 & 0.89 & 17.52$\pm$0.12 & >0.6 & 21.882(6) & >4.4 & W324 & No\\
        SW4 & 0.06 & 15.81$\pm$0.10 & >0.4 & 19.74(4) & >3.9 & W30 & Yes\\
        \hline
        \hline
    \end{tabular}
    \caption{Summary of the parameters of our four UV transients. The distances to the cluster centre are those given by \citetalias{Rivera_2018} (which they determined using the coordinates of the centre of 47 Tuc determined by \citealp{Goldsbury_2010}) for their sources with which our UV transients are associated. The observed outburst amplitudes (taken from the brightest observed \textit{uvm2} magnitude compared with the cluster background brightness at the source location) are lower limits because the true \textit{uvm2} quiescent magnitude of our transients could not be measured owing to the high cluster background, and because the outbursts are unlikely to be caught at maximum brightness. To determine a better constraint for the outburst amplitudes, we compared the \textit{uvm2} detections with the U300 magnitude measured by \citetalias{Rivera_2018} (using HST) from the most likely quiescent counterpart of our sources, although these are still lower limits as the transients may not have been observed at during the peak of their outbursts in the UVOT observations. Photometric errors for the U300 magnitude (taken from \citetalias{Rivera_2018}) values are given in parentheses.}
    \label{tab:sum_results}
\end{table*}

We detected four UV transients (and one UV variable, which could be identified as an RR Lyrae star; see Section 3.2.6) in the 75 \textit{uvm2} observations obtained with the UVOT of the Galactic GC 47 Tuc. All four transients are likely DN outbursts from CVs, of which only one has previously been observed in outburst (see Table~\ref{tab:sum_results} for details). We thus detected three new GC DNe candidates, of which there are, as of this study, only 17 known in all Galactic GCs. These identifications are primarily based on the outburst properties and the positional coincidence of our UV sources with known CVs or candidate CVs in 47 Tuc as identified by \citetalias{Rivera_2018} (and previously \citealp{Edmonds_2003}).\

The observed changes in \textit{uvm2} magnitude above the cluster background that we detected during the brightest outbursts were 1.06, 0.67, 0.59, and 0.41 for SW1--4, respectively. However, if we take the values for the HST U300 magnitude of the CV or CV candidate counterparts to our sources as estimates for the brightness of our sources in quiescence, we obtain for SW1--4 a change on the magnitude scale of at least 4.1, 3.2, 4.3, and 3.9, respectively. These all lie within the 2--6 mag range of expected (optical) brightenings of DNe (see e.g. \citealp{Dobrotka_2006,Belloni_2019}). These outburst amplitudes are lower limits, as outbursts are not necessarily caught at maxima during the UVOT observations.\

The sources SW1 and SW3 have not been observed in outburst before. We detected five outbursts from SW1 and one from SW3. \citet{Edmonds_2003} suggested that the source corresponding to SW2 may have exhibited `signs of outbursts' owing to its strong variability, although they could not conclusively identify the variability as being caused by DN outbursts. We detected several clear outbursts of this source, confirming its DN nature. The location of SW4 corresponds to the known DN V2, which has been observed in outburst twice \citep{Paresce_1994,Shara_1996}. We detected four additional outbursts of this source. Until now it was one of only two known DNe in 47 Tuc; the second is AKO 9, which was not detected in the UVOT observations. We therefore increased the DN sample in this cluster to five.\ 

As described in Sections 3.2.1--3.2.4, the observational sampling and high cluster background for the observations of 47 Tuc made obtaining estimates for durations and recurrence times for the outbursts difficult. Nevertheless, our light curves show that the outburst durations for all four sources occurred on timescales of days to weeks, as expected from DNe. We obtained estimates for the duty cycles (the fraction of time a source is in outburst) of our sources, which is equal to the ratio of the average outburst duration time to the average recurrence time (both measured over a time span covering many outbursts). We used the upper limits for the duration times and the shortest observed recurrence times (or lower limit in the case of SW3) to compute duty cycle upper limits for each source, yielding <0.1, <0.3, <0.5, and <0.1 for SW1--4, respectively. Duty cycles can also be obtained as the ratio of the number of outburst detections to the total number of observations, for which we find consistent results with duty cycle values of 0.08, 0.22, 0.04, and 0.07 for SW1--4, respectively. These values are consistent with typical DNe: for instance, \citet{Coppejans_2016} found DN duty cycles ranging from 0.01-0.36, using light curves spanning 8 to 9 years of several hundred events from the Catalina Real-Time Transient Survey.\

We also note that all the values we infer for the \textit{uvm2} luminosities of SW1--4 (see Table~\ref{tab:fluxes_lums}) lie within the range of typical UV luminosities exhibited by DN (${\sim}10^{33}-10^{35}$ erg s$^{-1}$; \citealp{Wheatley_2000} and references therein; \citealp{Ramsay_2010}).\ 

Simultaneous XRT observations revealed no X-ray sources at the locations of SW1, SW2, and SW3 during outburst (see Table~\ref{tab:fluxes_lums} for upper limits on the X-ray fluxes and luminosities). Typical X-ray luminosities for DNe in the 0.5--10 keV band range ${\sim}10^{30}$--$10^{34}$erg s$^{-1}$ \citep{Balman_2015,Wada_2017}. The upper limits we determined for SW1--3 are consistent with this range and with the Chandra (0.5-6 keV) luminosities derived by \citet{Heinke_2005}. However, this result should be taken with caution, as X-ray emission from DNe may lag with respect to the optical and UV radiation and X-rays may decrease in brightness during the DN outbursts \citep{Wheatley_2003}. We could not derive any estimates or upper limits of the X-ray emission from SW4 because its location in the XRT images is significantly contaminated by strong X-ray flux from the nearby source X9, a well-known X-ray binary (see e.g. \citealp{Heinke_2005,Bahramian_2017}). Additionally, inspecting the XRT images corresponding to both observations during the UV outbursts and when the source was not in outburst revealed no change in the X-ray flux at the location of SW4 (within the errors on the fluxes).\

We caution that the association of our transients with the known CVs or CV candidates is not entirely conclusive, since multiple HST sources are present within the positional error circle of our UV transients. However, the identification of our sources as counterparts to the known CV or CV candidates is strengthened by the fact that for all sources there is only one CV or CV candidate (from \citetalias{Rivera_2018}) within the positional error circle, and that the outburst properties they exhibit are clear signatures of DN outbursts.\

Until now only 14 DNe have been detected in GCs. With the sources SW1, SW2, and SW3, we increase this number significantly to 17 and we increase the total number of DNe in 47 Tuc from 2 to 5, making it the GC with the highest number of observed DNe (no other GC contains more than 2 identified DNe, see \citealp{Pietrukowicz_2008} and \citealp{Belloni_2016}). This low number of detections is in stark contrast with theory, which predicts that CVs should be very common in GCs (>100 in massive GCs) and that around 50\% of CVs should exhibit DN outbursts \citep{Downes_2001,Dobrotka_2006,Knigge_2011,Belloni_2016}. This discrepancy has been examined over multiple decades, and several possible explanations have been explored (see Section 1 for more information). Several studies have attributed the lack of observed events to an intrinsic difference in the physical properties of CVs in GCs compared with field CVs (e.g. \citealp{Dobrotka_2006}; see Section 1), while others have interpreted the problem as a result of observational constraints, in particular high GC background and stellar crowding (e.g. \citealp{Belloni_2016}; see Section 1). There have been some attempts to compute the expected observable DNe rates in GCs to quantify the discrepancy. \citet{Belloni_2016} model a theoretical massive cluster containing 108 CVs, and from typical DNe duty cycles determine the probability of detecting a CV (in quiescence or outburst) based on the distance to the cluster and the limiting (optical) magnitude of the observations (see Fig. 5 in their paper). Although the study concerns optical observations, to obtain an estimate from our data we can input values pertinent to the UVOT observations of 47 Tuc to determine that $\sim$2.5\% of CVs (2-3 for a cluster with 108 systems) should be detectable in outburst during any given (UVOT) observation. \citet{Dobrotka_2006} used similar methods/models to expect 0.5-2 DNe to be occurring in 47 Tuc at any given time. In our study, we have 31 detections during outburst from a total of 75 observations, indicating a rate of 0.41 detections per observation. This therefore suggests only a small discrepancy with the (lower estimates of) the expected rates. Considering this, and taking into account the uncertainties and assumptions associated with the theory used to make these predictions, the results of this study indicate that the discrepancy between theory and observations of DNe in GCs can likely be explained primarily by observational biases, and not by differences in the real rates of of GC DNe outbursts compared with DN outbursts from field CVs.\ 

We note that as a result of the very high cluster background emission at the core of the cluster, relatively weak outbursts may not appear clearly in the difference images and thus may not be picked up by our pipeline. For instance, most of the flux measurements from the outbursts of SW2 and SW3 are less than $1\times10^{-15}$ erg s$^{-1}$ cm$^{-2}\ \si{\angstrom}^{-1}$ above their respective backgrounds. If  these sources had been located at the position of SW4, we likely would not have detected them at all above the stronger background (and, consequently, larger error bars on the flux measurements; see Fig.~\ref{fig:lightcurves}). We can therefore deduce that from UVOT observations it is typically only possible to detect the very brightest DNe in the centre of the cluster. Since the number of interacting binaries is predicted to increase towards the cluster core \citep{Sigurdsson_1995,Benacquista_2013, vandenBerg_2019}, this suggests that many GC DNe will be not be detectable with UVOT observations.\ 

Stellar crowding is closely linked with the issue of background flux (both indicating high stellar densities) and can help explain the lack of detections. In optical and NUV studies of GCs with HST, higher resolution allows many more stars to be resolved, significantly decreasing their contribution to cluster background emission. On the other hand, the optical flux from GCs is significantly higher than the UV flux owing to typically old and thus UV-faint populations of stars dominating GCs; thus the optical cluster background emission may still be more significant than the UV for ground-based (and typically seeing limited) optical telescopes. Therefore, optical searches can be hindered by both the background flux and the lower optical amplitude of DNe outbursts with respect to the UV, suggesting that the lack of UV studies could be potentially contributing to the paucity of observed GC DNe. \ 

In searches for DNe in GCs, it is thus difficult to evaluate the benefits and challenges for each wavelength regime. Nevertheless, if we take note of the relative success of this UV study in corroborating theory with observations, we suggest that UV searches of GC DNe (and UV variability studies of GC in general) can be highly beneficial, in particular if higher angular resolutions can be achieved than what is possible with UVOT.

\subsubsection*{Acknowledgements}
The authors would like to thank the referee Paul Kuin for his insightful and useful comments to improve our paper. DM is partly supported by the Netherlands Research School for Astronomy (NOVA). ASP and RW acknowledge support from a NWO Top Grant, Module 1, awarded to RW.

\newpage

\bibliography{mybibliography}

\end{document}